\documentclass[sigconf]{acmart}

\usepackage{chngcntr}
\usepackage{graphicx}
\usepackage{amsmath}
\usepackage{subcaption}
\usepackage{amsfonts}
\usepackage{algorithmic}
\usepackage{soul}
\usepackage{url}
\usepackage{balance}
\usepackage{multirow, multicol, makecell}
\usepackage{url} 
\usepackage{xspace}
\usepackage{framed}

\usepackage{enumitem} 
\usepackage{xcolor}

\newcommand{\name}[1]{\textsc{SocratiCode}}

\title{Towards \name{}: Designing a Generative AI-Based Programming Tutor for K-12 Students through a 4-Week Participatory Design Study}

\begin{abstract}

Generative AI creates new opportunities for programming education, but many existing systems remain overly directive, producing lengthy explanations and premature solutions that can overwhelm K-12 novices. In this paper, we present a participatory design study of how an adaptive tutorial system, \name{}, evolved toward a Socratic tutoring model for beginner programming instruction. Drawing on weekly learner feedback, we iteratively refined the system over a four-week study with two K-12 students learning Python. Across iterations, the system shifted from flexible tutorial generation toward a more dialogic form of support characterized by guided questioning, reflection prompts, misconception checks, incremental hints, and mandatory pauses for learner input. Our preliminary observations suggest that this Socratic shift improved explanation clarity, supported problem-solving engagement, and better aligned instruction with novice learners’ needs, especially when combined with human guidance. We argue that generative AI in K-12 programming education may be most effective not as an answer engine, but as a Socratic, adaptive learning companion embedded within a human-guided instructional framework.

\end{abstract}



\copyrightyear{2026}
\acmYear{2026}
\setcopyright{cc}
\setcctype{by}

\author{Cassandra Lucas}
\affiliation{%
 \institution{Harry A. Burke High School}
  \city{Omaha, Nebraska}
 \country{USA}
}
\email{cassiebobassy2020@gmail.com}

\author{Anshul Bihani}
\affiliation{%
 \institution{Millard North High School}
  \city{Omaha, Nebraska}
 \country{USA}
}
\email{anshulbihani5@gmail.com}

\author{Rohini Kukka}
\affiliation{%
 \institution{Missouri University of Science and Technology}
  \city{Rolla, Missouri}
 \country{USA}
}
\email{rkqfg@mst.edu}

\author{Chun-Hua Tsai}
\affiliation{%
 \institution{University of Nebraska  Omaha}
 \city{Omaha, Nebraska}
 \country{USA}
}
\email{chunhuatsai@unomaha.edu}

\author{Jaydeb Sarker}
\affiliation{%
 \institution{University of Nebraska  Omaha}
 \city{Omaha, Nebraska}
 \country{USA}
}
\email{jsarker@unomaha.edu}

\author{Mia Mohammad Imran}
\affiliation{%
 \institution{Missouri University of Science and Technology}
  \city{Rolla, Missouri}
 \country{USA}
}
\email{imranm@mst.edu}

\begin{document}
\maketitle

\begin{sloppypar}

\section{Introduction}
\label{sec:intro}
Generative AI has recently emerged as a powerful tool for content creation in education~\cite{DepartmentForEducation2025, mittal2024comprehensive, gu2025ai}, including programming instruction. Large language models (LLMs) such as ChatGPT and Claude can generate explanations, examples, and step-by-step tutorials on demand, creating new opportunities for scalable and personalized learning support. These capabilities hold particular promise for programming education~\cite{DepartmentForEducation2025, mittal2024comprehensive}, where learners often benefit from immediate explanations, worked examples, and interactive assistance while developing conceptual understanding and problem-solving skills.


Despite this promise, many generative AI systems remain fundamentally answer-oriented. They can produce tutorial-like content quickly, but often struggle to engage learners in the reasoning processes needed for durable understanding. First, such systems often lack mechanisms to determine appropriate stopping points, resulting in overly long or unfocused content~\cite{giannakos2025promise}. Second, they are typically unaware of the learner’s prior knowledge or expertise level, which can lead to mismatches between instructional content and learner needs~\cite{tasdelen2025generative}. Third, they may introduce concepts that are unfamiliar, irrelevant, or out of scope~\cite{DepartmentForEducation2025}. Finally, they may generate content that is misaligned with the learner’s goals, thereby creating confusion rather than supporting understanding~\cite{giannakos2025promise, gu2025ai, klopfer2024generative}. In programming education, these limitations are especially problematic because novices need not only explanations but also opportunities to articulate their reasoning, reflect on misunderstandings, and work through problems before receiving complete solutions.

These challenges can be particularly pronounced in K--12 programming education, where many learners have little prior exposure. Compared to older students, K--12 learners face greater difficulties with abstract reasoning, pacing, and engagement~\cite{grover2013computational, kelleher2005lowering}, which underscores the importance of scaffolding, structured guidance, and developmental alignment in computational thinking education~\cite{grover2013computational}. Research has shown that younger learners benefit from instructional approaches that incorporate storytelling, real-world analogies, and stepwise pacing, as these strategies help connect abstract code to familiar experiences~\cite{kelleher2005lowering, reiser2018scaffolding}. Without careful regulation and design, generative AI can overwhelm K--12 students with content that is too advanced, too lengthy, or insufficiently contextualized, thereby undermining effective learning principles that emphasize concise structure, reinforcement, and gradual progression~\cite{winslow1996programming}.


Motivated by a need to deeply understand this problem space, we adopt an exploratory participatory design approach to investigate how an adaptive tutorial system might support a more dialogic and learner-responsive form of instruction.
Rather than treating generative AI only as a mechanism for tutorial delivery, we investigate how it can support a Socratic tutoring model. The Socratic method is a teaching approach in which learning is guided through structured questions that prompt learners to explain their reasoning, reflect on assumptions, identify misunderstandings, and work toward answers on their own rather than receiving them directly~\cite{alshaikh2020socratic, alshaikh2020experiments, al2023socratic, britannica_socratic_method}.

This study emphasizes guided questioning, reflection, incremental hints, and learner-paced progression~\cite{alshaikh2020socratic, alshaikh2020experiments, al2023socratic, britannica_socratic_method}. In computational tutoring systems, researchers operationalize these principles through targeted follow-up questions that elicit student predictions, probe explanations, and surface misconceptions based on the learner's actual responses, prioritizing reasoning over direct answer delivery~\cite{alshaikh2020socratic, alshaikh2020experiments}. We adopt this computational interpretation rather than the classical dialectical form.

We refer to the proposed framework as \name{}. The framework draws on adaptive learning theory~\cite{brusilovsky2007user}, research on generative AI in education~\cite{brachman2025building, urhan2024problem}, recent advances in prompt engineering~\cite{drosos2025dynamic, white2023prompt}, and Socratic tutoring principles that emphasize guided questioning, reflection, and incremental support~\cite{alshaikh2020socratic, alshaikh2020experiments, al2023socratic, britannica_socratic_method}. We initially developed the system as an adaptive, prompt-based tutorial generator and subsequently refined it to better support learner reasoning through structural constraints, controlled pacing, misconception clarification, opportunities for learner response, and question-driven interaction.

More specifically, \name{} mitigates overly long or unfocused tutorials by embedding structural elements such as introductions, examples, practice, summaries, and follow-up tasks that maintain coherence and create natural stopping points~\cite{white2023prompt}. The system also adapts to individual learners by eliciting minimal background information and adjusting pacing, analogies, and explanations to provide more developmentally appropriate support in K--12 settings~\cite{park2025generative}. As the design evolved, we incorporated additional scaffolding strategies, including misconception clarification, metaphor use, reflective pauses, and hints provided before complete solutions~\cite{park2025generative, white2023prompt}. Through these refinements, the system moved from content delivery toward guided inquiry.

We developed this framework through a participatory design process with two K--12 students, who engaged with the adaptive prompt-based system on the OpenAI platform over a four-week period in Summer 2025. We iteratively refined the prompt based on weekly learner feedback to better align explanations, pacing, interaction style, and instructional structure with their developmental needs. This process showed how learner feedback pushed the system away from flexible tutorial generation and toward a more controlled tutoring style centered on questioning, reflection, and staged support. Our preliminary findings suggest that adaptive prompting, combined with human oversight, can help generative AI function more effectively as a Socratic learning companion for beginner programmers. Our study has following contributions:

\begin{itemize}[leftmargin=*]
    \item We present a participatory design study of how an adaptive prompt-based generative AI system evolved toward a Socratic tutoring model for K--12 programming learners.
    \item We design and iteratively refine \name{}, a learner-sensitive framework that integrates pedagogical scaffolding, guided interaction, and adaptive support for novice programmers.
    \item We provide preliminary evidence that iterative prompt refinement, informed by learner feedback, can shift generative AI from tutorial-style explanation delivery toward a more dialogic form of support grounded in questioning, reflection, and human guidance.
    \item We provide the full prompt specification in a replication package to support future research~\cite{anonymous_2025_17220238}.
\end{itemize}

\section{Background and Related Work}
\label{sec:background}

Recent studies show that novice programmers often become over-reliant on generative AI tools. Rahe and Maalej found that students repeatedly used ChatGPT to generate complete solutions after initial failures, which reduced persistence and led to inefficient debugging~\cite{rahe2025programming}. Choudhuri \textit{et al.}~\cite{choudhuri2025utilize} similarly found that while AI can support coding tasks, it often hallucinates, provides weak rationales, and fails to adapt to learners’ needs, leaving students confused and overly dependent on external solutions. Zi \textit{et al.}~\cite{zi2025struggle} further showed that beginners struggle both to craft prompts and to interpret AI-generated code, often overestimating its correctness. Nguyen \textit{et al.} argue that novices must actively debug, question, and evaluate generated solutions rather than use AI passively~\cite{nguyen2024misread}. In K--12 education, where generative AI is increasingly being introduced~\cite{gu2025ai}, researchers have called for careful experimentation and pedagogically grounded adaptation~\cite{klopfer2024generative}. Therefore,  integration of generative AI systems for K--12 should support scaffolding, debugging, and active reasoning rather than simply produce code or tutorial-style explanations.

This need aligns with the pedagogical logic of Socratic instruction. The Socratic method emphasizes guided questioning and learner reflection rather than direct answer delivery~\cite{britannica_socratic_method}. This perspective is also relevant to programming learning, where Socratic tutoring has been used to help novices explain code, predict behavior, and refine understanding through guided questions instead of immediate solutions~\cite{alshaikh2020socratic, alshaikh2020experiments, al2023socratic}. A Socratic tutoring model, therefore, offers a useful lens for designing generative AI systems that support learner reasoning, misconception repair, and incremental guidance in beginner programming contexts.

To inform the initial prompt design, we reviewed prior research on adaptive learning and programming pedagogy~\cite{brusilovsky2007user, reiser2018scaffolding, kelleher2005lowering}. Prior work highlights storytelling, real-world analogies, and scaffolded pacing as effective supports for understanding complex concepts~\cite{kelleher2007storytelling, suh2022codetoon, winslow1996programming, alasmari2023current}. The literature also emphasizes clarifying misconceptions~\cite{narciss2025learning}, using metaphors to connect abstract ideas to familiar contexts~\cite{suh2022codetoon}, reinforcing lessons through summaries~\cite{palinscar1984reciprocal}, and providing follow-up problems to support continued engagement and progressive learning~\cite{denny2024explaining}. These strategies are consistent with both adaptive learning and a Socratic tutoring perspective centered on questioning, reflection, and structured support.

Our work builds on these strands of research by examining how a prompt-based generative AI system can be iteratively shaped, through participatory design, from an adaptive tutorial generator into a more dialogic and learner-responsive tutoring model. Rather than treating adaptivity only as personalization of content delivery, we investigate how it can also support Socratic features such as paced interaction, guided reflection, misconception-oriented feedback, and hints before full solutions.

\section{Methodology}
\label{sec:method}

\begin{figure*}[t]
    \centering
    \includegraphics[width=0.8\linewidth]{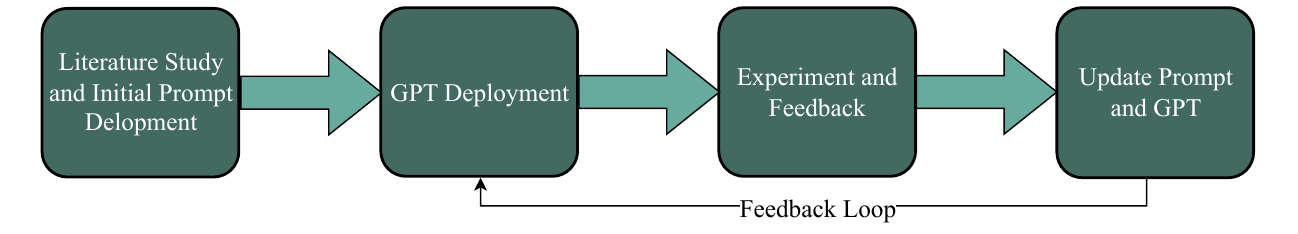}
    \caption{Experiment Pipeline of \name{}.}
    \label{fig:experiment-pipeline}
\end{figure*}

\subsection{\name{} Design}

The initial prompt for \name{} was designed based on prior work on auto-tutors, programming tutorials, and recent studies on generative AI for content creation and programming assistance~\cite{d2013autotutor, winslow1996programming, almaiah2022examining, brachman2025building, urhan2024problem, fan2025impact, boguslawski2025programming}. In particular, the guidelines of Winslow \textit{et al.}~\cite{winslow1996programming}, Brusilovsky \textit{et al.}~\cite{brusilovsky2007user}, and Boguslawski \textit{et al.}~\cite{boguslawski2025programming} informed the initial draft.

To ensure adaptivity, the model was instructed to collect background information from learners prior to generating tutorial content:
\textit{“What is your background? Please select the option that best describes you: (1) I have never engaged in related activities before; (2) I have watched a few tutorials or videos; [...]”}
Tutorials were then generated to align with the learner’s self-reported experience and background. Since this study focused exclusively on beginners, we instructed the model to assume the user was always a beginner. To ensure consistency, all tutorials were generated in Python unless learners explicitly requested an alternative programming language.

The prompt was structured into multiple instructional components. At a high level, it comprised the following: System Role and General Instructions; Learner Level and Background Selection; Tutorial Structure and Flow Control; Reinforcement, Adaptivity, and Closure; and Constraints and Content Boundaries. Two authors collaboratively drafted the initial version of the prompt, grounding it in prior literature. The initial design primarily functioned as an adaptive explanatory tutor. However, as the study progressed, the prompt was iteratively refined through learner feedback toward a more dialogic and learner-responsive tutoring model with increasingly Socratic features, including guided interaction, reflection, paced progression, and delayed answer delivery.

\subsection{Experiment Pipeline \& K--12 Participants}

Figure~\ref{fig:experiment-pipeline} illustrates the overall experimental pipeline. The first version of the \name{} prompt was developed through literature-informed design by the authors and subsequently deployed on the \textit{GPT} platform to ensure accessibility for learners. We chose GPT-5 as our default model. We then designed a four-week curriculum in which two K--12 students engaged with programming topics through the adaptive tutorials. Participants provided both daily and weekly feedback on their learning experiences. Based on this feedback, the prompt was revised at the beginning of the following week when necessary.

We devised a four-week curriculum covering fundamental programming topics aligned with the standard ACM/IEEE Computer Science Curricula (CS2013) introductory programming guidelines~\cite{CS2013}. During the study, a teaching assistant (TA), who is a master's student in computer science, introduced a new topic and provided 3--4 practice problems designed to reinforce the concept. The problems were selected to ensure a balance between conceptual understanding and hands-on application, and participants were encouraged to attempt them independently before consulting the adaptive tutorials. In weekly meetings, a computer science faculty member conducted one-to-one and group meetings with participants. Moreover, for each topic, we provided the initial prompt, and the details are available in the replication package~\cite{anonymous_2025_17220238}.

During the summer of 2025, the university recruited local high school interns to participate in research projects. Two of these interns, one male and one female, both in grade 11, joined our team for six weeks. They were between 17 and 18 years old and had no prior programming experience. Upon submission of the application to the Institutional Review Board, we received approval for this study, and we will share approval documentation upon acceptance of the paper. The study itself was conducted over four weeks, corresponding to weeks 2--5 of the internship. Week 1 was reserved for logistical preparation, and week 6 for final wrap-up activities. For clarity, we refer to weeks 2--5 as W1--W4 throughout the remainder of the paper.

\subsection{Participatory Prompt Design}

\begin{figure*}[t]
    \centering
    \includegraphics[width=0.8\linewidth]{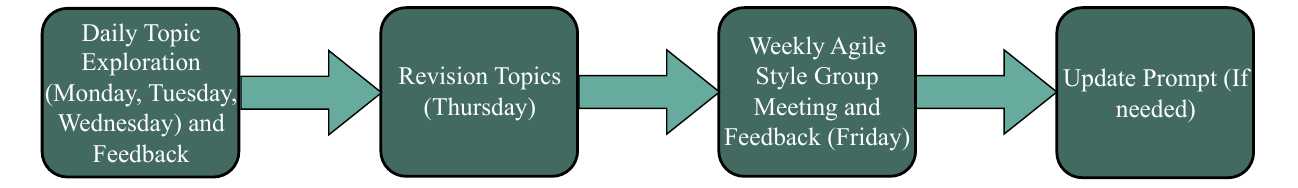}
    \caption{Feedback Loop}
    \label{fig:feedback-loop}
\end{figure*}

Because the initial prompt had previously been refined with feedback from undergraduate students who already possessed some programming knowledge, we anticipated that it might not fully meet the needs of younger or less experienced learners in K--12 education. To address this limitation, we adopted an iterative participatory design process during the four-week study, in which participants provided both daily and weekly feedback on their tutorial experiences. Prompt revisions were introduced at the beginning of W2, and the designs used in W3 and W4 were informed by feedback collected during prior weeks. No further modifications were necessary in W4, indicating that the prompt had stabilized. The overall procedure for collecting feedback and updating prompts is illustrated in Figure~\ref{fig:feedback-loop}.

Across W1--W4, the system evolved from adaptive explanatory tutorialing toward a more dialogic tutoring flow with increasingly Socratic characteristics. In W1, the prompt emphasized adaptive explanatory tutorialing by defining the core role, learner profile, and tutorial structure, while still allowing forward-looking guidance and relatively soft constraints. In W2, the prompt introduced stronger pacing and more controlled learner interaction, reducing premature topic expansion and improving instructional consistency. In W3, the design further incorporated explicit checkpoints, construct isolation, and increased use of guided hints, encouraging learners to engage with one concept at a time before moving forward. By W4, the prompt had stabilized into a structured interactive flow characterized by pauses, reflection, learner response requirements, and hard stopping points. At this stage, the system no longer suggested unsolicited next steps and instead ended lessons explicitly, for example, replacing \textit{“Next: Let’s talk about decisions with conditions”} with \textit{“This completes the lesson.”} These revisions produced a more structured and dialogic tutoring flow.

We provide the shortened prompt template used by the end of W4 below. 


\smallskip
\begin{promptbox}{Template Structure (Shortened). For Full Prompt~\cite{anonymous_2025_17220238}}
{
\textbf{1. Role \& Audience}:  
Act as a step-by-step tutorial guide for absolute beginners, using Python by default and clear analogies.

\textbf{2. Learner Adaptation}:  
Ask for the learner profile first and tailor explanations accordingly.

\textbf{3. Lesson Flow}:  
Hook or analogy, concept explanation, code walkthrough, short exercise, optional misconception note, reflection, transition.

\textbf{4. Reinforcement}:  
Use checkpoints and hints; increase difficulty gradually with explicit guidance.

\textbf{5. Tutorial Closure}:  
End with a brief summary and one challenge, then \textit{hard-stop}.

\textbf{6. Interaction Rules}:  
Pause after exercises and reflections; give hints before solutions; proceed only after learner input.

\textbf{7. Scope \& Constraints}:  
Teach only requested concepts, omit extras, maintain statelessness, and use only introduced constructs.
}
\end{promptbox}

\subsection{Data Collection \& Analysis}

We designed a mixed-methods protocol inspired by agile practices, in which daily stand-ups and weekly syncs incorporated qualitative feedback through open-ended questions and quantitative feedback through surveys. All surveys were administered via Google Forms, while reflections and interviews conducted during daily stand-ups and weekly sync meetings took place in person or over Zoom. These data were used to examine how beginners learned with generative AI and to support reflection on participatory prompt design practices.

\noindent
\textbf{Daily Stand-Ups.} The daily reflection questions included:

\begin{itemize}[leftmargin=*]
    \item \textbf{Closed-ended:}
    \begin{itemize}[leftmargin=*]
        \item Did you complete the task? (Yes / No / Partially)
        \item Have you worked on this type of problem before? (Yes / No)
        \item Did you consult an expert for help? (Yes / No)
    \end{itemize}

    \item \textbf{Open-ended:}
    \begin{itemize}[leftmargin=*]
        \item What explanation did the expert provide?
        \item What did you understand well today?
        \item What confused you today, if anything?
        \item Do you have suggestions to improve the adaptive tutorial?
        \item Any additional thoughts on improving your learning with generative AI?
    \end{itemize}
\end{itemize}

\noindent
\textbf{Weekly Sync Meetings.} The weekly reflection questions included:

\begin{itemize}[leftmargin=*]
    \item \textbf{Weekly survey questions:}
    \begin{itemize}[leftmargin=*]
        \item Was the adaptive version more helpful than earlier in the week? (from W2 onward) (Yes / No)

        \item \textbf{5-point Likert scale:}
        \begin{itemize}[leftmargin=*]
            \item The tutorial matched my current skill level.
            \item Explanations and examples were clear and concise.
            \item The tutorial adapted to my learning difficulties.
            \item The structure (checkpoints, examples, summaries) supported my understanding.
        \end{itemize}

        \item \textbf{Open-ended:}
        \begin{itemize}[leftmargin=*]
            \item Which tutorial elements were most helpful?
            \item Additional suggestions for improving the adaptive tutorial?
        \end{itemize}
    \end{itemize}
\end{itemize}

Two authors independently conducted open coding and thematic analysis on weekly reflections, daily surveys, and observational notes to explore how adaptive prompting and human oversight can support K--12 learning. Using an inductive approach~\cite{charmaz2006constructing}, we generated initial codes through iterative review, then refined them through discussion and disagreement resolution. Four key categories emerged: \textit{Engagement and Appeal}, \textit{Human--AI Collaboration}, \textit{Explanations and Clarity}, and \textit{Instructional Design and Structure}. This thematic framework captures recurring patterns in learner feedback and interactions throughout the W1--W4 participatory study and helps explain how prompt refinement shaped the K--12 learning experience.

\section{Findings}
Based on daily and weekly feedback, we conducted a preliminary thematic analysis to report early findings and observations.

\subsection{\textbf{Theme 1: Guided Questioning Improved Learner Engagement}}

\smallskip
\noindent
\textbf{Observation 1.1: Engagement was strengthened when the system adapted to learners' pace and reasoning.}
Participants reported that \name{} was engaging, particularly as a support tool for novice learners. This engagement was attributed not only to the novelty of generative AI, but also to the system’s ability to adapt explanations to learners’ prior knowledge, pace, and responses. Both participants noted that the system provided relevant examples that supported understanding. P1 stated, \textit{``I enjoyed how AI adapted to fit my learning style. It built off what I already knew and gave many relative anecdotes and examples to help me learn.''} Weekly reflections further indicated that engagement was sustained as learners progressed, with W4 noting, \textit{“Learning the concepts from the \name{} went really well; I understand the basics of each lesson I went over with the GPT and how to apply them to simple programming problems.”}

\smallskip
\noindent
\textbf{Observation 1.2: Engagement depended on explanations that supported active reasoning rather than passive answer reception.}
Participants emphasized that engagement with programming concepts was closely tied to how clearly and interactively explanations were presented. While the initial versions of the system did not always provide sufficient support for beginners, later iterations were perceived as clearer and more effective. Daily feedback suggested that explanations became increasingly understandable as the prompt was refined, particularly when the system slowed its pace, used relevant examples, and better aligned its responses with learners' reasoning process.

\subsection{\textbf{Theme 2: Reflection and Checkpoints Supported Understanding}}

\smallskip
\noindent
\textbf{Observation 2.1: Definitions, checkpoints, and pacing improved conceptual clarity.}
Although participants generally described the tutorial explanations as clear, clarity decreased when new topics were introduced without first defining key terms or checking for understanding. Early iterations occasionally moved through multiple ideas without sufficient pause, which overwhelmed novices unfamiliar with programming vocabulary. Participants recommended brief definitions, staged progression, and explicit checks for understanding before advancing. Both participants noted that later adjustments, especially the addition of checkpoints, helped mitigate these issues.

\smallskip
\noindent
\textbf{Observation 2.2: Breaking concepts into smaller steps and linking them to familiar contexts supported learning.}
Participants reported that breaking programming concepts into smaller modules and allowing them to move forward only after understanding earlier material improved comprehension. Connecting abstract programming ideas to real-world analogies further supported understanding and made the lessons feel more accessible. Learners also valued structured explanations, examples, and reflection opportunities as part of this more gradual instructional flow.

\subsection{\textbf{Theme 3: Incremental Hints Were Preferred Over Immediate Solutions}}

\smallskip
\noindent
\textbf{Observation 3.1: Learners valued guided hints and multiple attempts before full answers.}
Participants consistently preferred support that preserved opportunities for problem solving rather than immediately revealing solutions. Structured practice, examples, and step-by-step walkthroughs were useful, but learners specifically highlighted the value of being able to try problems on their own and receive hints when needed. One participant stated, \textit{“The GPT also asked me if I wanted a hint or if I wanted to do the problem on my own… which was a nice option.”} Early in the study, participants explicitly requested more guided hints instead of full solutions. P1 suggested, \textit{“Let the user try multiple times to answer a question… If they ask for a hint, provide a small but relative hint to help them along.”}

\smallskip
\noindent
\textbf{Observation 3.2: Immediate or inconsistent explanations sometimes created confusion.}
Participants reported confusion when new syntax or code constructs appeared without sufficient explanation. P1 commented that the system \textit{“provided many examples, and explained them quite well,”} but also described moments when explanations were missing. P2 noted that similar problems were sometimes addressed using different solution styles across users, which made comparison less straightforward. As the study progressed, confusion decreased, but it remained more likely when the system moved too quickly, introduced unfamiliar material, or provided answers without enough intermediate support.

\subsection{\textbf{Theme 4: Human Oversight Remained Necessary for Deeper Guidance}}

\smallskip
\noindent
\textbf{Observation 4.1: Human assistance remained essential alongside AI-based tutoring.}
Both participants emphasized that human guidance remained important for effective programming learning alongside \name{}. P1 noted, \textit{“I do think that expert assistance will always be needed, especially in coding…”}, while P2 suggested that \textit{“two to three times human assistance per week would be enough.”} Human assistance helped clarify challenges not fully covered by the tutorial and supported learning when students encountered more advanced or context-dependent material.

\smallskip
\noindent
\textbf{Observation 4.2: Human support was especially valuable for misconceptions and advanced topics.}
Weekly reflections indicated gradual progression from variables and conditionals in W1 to loops, arrays, and functions by W4. P1 described gaining clarity around specific misconceptions, stating, \textit{“I understood Out-of-Bounds Access [...] forgetting that in Python indexes start at 0, and mistaking conditional ‘=’ with comparing ‘==’.”} While participants valued the tutorial’s availability and responsiveness, they noted limitations when working without expert input. Midway through the study, both participants also noted that \name{} sometimes acted as a “yes-man,” prioritizing its own answers over learner reasoning. Occasional human intervention was therefore particularly valuable for correcting misconceptions, challenging weak reasoning, and supporting more advanced topics such as recursion and functions.

\section{Conclusion, Limitations \& Future Work}
\label{sec:conclusion}

We conducted a four-week participatory design study with two K--12 students to examine how \name{} could support beginner programming learning. Across weekly revisions, the system evolved from adaptive tutorial generation toward a more Socratic tutoring model shaped by participant feedback. In particular, the prompt increasingly emphasized questioning, reflection, paced progression, and guided hints before full solutions. Our findings suggest that adaptive generative AI, when combined with human guidance, can support K--12 programming education by improving clarity, structure, and learner engagement.

Our findings further suggest that generative AI is most useful in K--12 programming education when designed as a learner-responsive instructional companion rather than a standalone answer generator. Participants benefited from clear explanations, examples, analogies, checkpoints, and guided hints, while human support remained important for addressing misconceptions, challenging weak reasoning, and helping with more advanced topics. These results highlight the value of designing AI systems that support Socratic forms of guidance while remaining embedded within a human-guided instructional framework.

This study has several limitations. First, we worked with only two K--12 participants, which limits the generalizability of our findings. Second, we used a single customized GPT-based system, and other generative AI models may behave differently under similar conditions. Third, our study focused only on Python, which may limit transferability to other programming languages. Finally, the underlying model may reflect biases in training data, representation, and instructional style that could influence learner experiences and outcomes.

In future work, we plan to evaluate these findings with larger and more diverse K--12 populations across grade levels, backgrounds, and prior programming experience. We also plan to examine how problem difficulty, pacing, and interaction design can be better aligned with learner progress over time. In particular, we aim to design adaptive generative AI systems that further strengthen Socratic support for beginner programmers through exploration, multiple attempts, reflection, and incremental hints before full solutions.

\bibliographystyle{ACM-Reference-Format}
\balance
\bibliography{references}

@article{rahe2025programming,
  title={How Do Programming Students Use Generative AI?},
  author={Rahe, Christian and Maalej, Walid},
  journal={Proceedings of the ACM on Software Engineering},
  number={FSE},
  year={2025},
  publisher={ACM New York, NY, USA}
}

@article{almaiah2022examining,
  title={Examining the impact of artificial intelligence and social and computer anxiety in e-learning settings: Students’ perceptions at the university level},
  author={Almaiah, Mohammed Amin and Alfaisal, Raghad and Salloum, Said A and Hajjej, Fahima and others},
  journal={Electronics},
  volume={11},
  number={22},
  pages={3662},
  year={2022},
  publisher={MDPI}
}

@article{grover2013computational,
  title={Computational thinking in K--12: A review of the state of the field},
  author={Grover, Shuchi and Pea, Roy},
  journal={Educational researcher},
  volume={42},
  number={1},
  pages={38--43},
  year={2013},
  publisher={Sage Publications Sage CA: Los Angeles, CA}
}

@article{park2025generative,
  title={Generative AI prompt engineering for educators: Practical strategies},
  author={Park, Jiyeon and Choo, Sam},
  journal={Journal of Special Education Technology},
  volume={40},
  number={3},
  pages={411--417},
  year={2025},
  publisher={SAGE Publications Sage CA: Los Angeles, CA}
}

@inproceedings{drosos2025dynamic,
  title={Dynamic Prompt Middleware: Contextual Prompt Refinement Controls for Comprehension Tasks},
  author={Drosos, Ian and Williams, Jack and Sarkar, Advait and Wilson, Nicholas and Rintel, Sean and Panda, Payod},
  booktitle={Proceedings of the 4th Annual Symposium on Human-Computer Interaction for Work},
  pages={1--23},
  year={2025}
}

@article{white2023prompt,
  title={A prompt pattern catalog to enhance prompt engineering with chatgpt},
  author={White, Jules and Fu, Quchen and Hays, Sam and Sandborn, Michael and Olea, Carlos and Gilbert, Henry and Elnashar, Ashraf and others},
  journal={arXiv preprint arXiv:2302.11382},
  year={2023}
}

@techreport{DepartmentForEducation2025,
  author       = {Department for Education},
  title        = {Generative artificial intelligence (AI) in education},
  institution  = {Department for Education, UK},
  year         = {2025},
  month        = {August},
  note         = {Updated 12 August 2025},
  howpublished = {Government policy paper},
}

@article{tasdelen2025generative,
  title={Generative AI in the classroom: Effects of context-personalized learning material and tasks on motivation and performance},
  author={Tasdelen, Osman and Bodemer, Daniel},
  journal={International Journal of Artificial Intelligence in Education},
  year={2025},
  publisher={Springer}
}

@article{giannakos2025promise,
  title={The promise and challenges of generative AI in education},
  author={Giannakos, Michail and Azevedo, Roger and others},
  journal={Behaviour \& Information Technology},
  year={2025},
  publisher={Taylor \& Francis}
}

@article{boguslawski2025programming,
  title={Programming education and learner motivation in the age of generative AI: student and educator perspectives},
  author={Boguslawski, Samuel and Deer, Rowan and Dawson, Mark G},
  journal={Information and Learning Sciences},
  year={2025},
  publisher={Emerald Publishing Limited}
}

@article{d2013autotutor,
  title={AutoTutor and affective AutoTutor: Learning by talking with cognitively and emotionally intelligent computers that talk back},
  author={D'mello, Sidney and Graesser, Art},
  journal={ACM Transactions on Interactive Intelligent Systems (TiiS)},
  year={2013},
}

@inproceedings{urhan2024problem,
  title={Problem-Solving Through Pair-Programming: The Mediational Role of ChatGPT},
  author={Urhan, Selin and Kocadere, Selay Arkun},
  booktitle={2024 5th International Conference in Electronic Engineering, Information Technology \& Education},
  year={2024},
  organization={IEEE}
}

@article{fan2025impact,
  title={The impact of AI-assisted pair programming on student motivation, programming anxiety, collaborative learning, and programming performance: a comparative study with traditional pair programming and individual approaches},
  author={Fan, Guangrui and Liu, Dandan and Zhang, Rui and Pan, Lihu},
  journal={International Journal of STEM Education},
  volume={12},
  number={1},
  pages={16},
  year={2025},
  publisher={Springer}
}

@inproceedings{brachman2025building,
  title={Building Appropriate Mental Models: What Users Know and Want to Know about an Agentic AI Chatbot},
  author={Brachman, Michelle and Kunde, Siya and Miller, Sarah and Fucs, Ana and Dempsey, Samantha and Jabbour, Jamie and Geyer, Werner},
  booktitle={Proceedings of the 30th International Conference on Intelligent User Interfaces},
  pages={247--264},
  year={2025}
}

@article{winslow1996programming,
  title={Programming pedagogy—a psychological overview},
  author={Winslow, Leon E},
  journal={ACM Sigcse Bulletin},
  volume={28},
  number={3},
  pages={17--22},
  year={1996},
  publisher={ACM New York, NY, USA}
}

@article{kelleher2005lowering,
  title={Lowering the barriers to programming: A taxonomy of programming environments and languages for novice programmers},
  author={Kelleher, Caitlin and Pausch, Randy},
  journal={ACM computing surveys (CSUR)},
  volume={37},
  number={2},
  pages={83--137},
  year={2005},
  publisher={ACM New York, NY, USA}
}

@incollection{reiser2018scaffolding,
  title={Scaffolding complex learning: The mechanisms of structuring and problematizing student work},
  author={Reiser, Brian J},
  booktitle={Scaffolding},
  pages={273--304},
  year={2018},
  publisher={Psychology Press}
}

@incollection{brusilovsky2007user,
  title={User models for adaptive hypermedia and adaptive educational systems},
  author={Brusilovsky, Peter and Mill{\'a}n, Eva},
  booktitle={The adaptive web: methods and strategies of web personalization},
  pages={3--53},
  year={2007},
  publisher={Springer}
}

@book{CS2013,
  title     = {Computer Science Curricula 2013: Curriculum Guidelines for Undergraduate Degree Programs in Computer Science},
  author    = {{Joint Task Force on Computing Curricula, Association for Computing Machinery (ACM) and IEEE Computer Society}},
  year      = {2013},
  publisher = {ACM Press and IEEE Computer Society Press},
  address   = {New York, NY, USA},
}

@incollection{denny2024explaining,
  title={Explaining code with a purpose: An integrated approach for developing code comprehension and prompting skills},
  author={Denny, Paul and Smith IV, David H and Fowler, Max and Prather, James and Becker, Brett A and Leinonen, Juho},
  booktitle={Proceedings of the 2024 on Innovation and Technology in Computer Science Education V. 1},
  pages={283--289},
  year={2024}
}

@inproceedings{suh2022codetoon,
  title={Codetoon: Story ideation, auto comic generation, and structure mapping for code-driven storytelling},
  author={Suh, Sangho and Zhao, Jian and Law, Edith},
  booktitle={Proceedings of the 35th Annual ACM Symposium on User Interface Software and Technology},
  year={2022}
}

@inproceedings{alasmari2023current,
  title={Do current online coding tutorial systems address novice programmer difficulties?},
  author={Alasmari, Ohud Abdullah and Singer, Jeremy and Bikanga Ada, Mireilla},
  booktitle={Proceedings of the 15th International Conference on Education Technology and Computers},
  pages={242--248},
  year={2023}
}

@inproceedings{kelleher2007storytelling,
  title={Storytelling alice motivates middle school girls to learn computer programming},
  author={Kelleher, Caitlin and Pausch, Randy and Kiesler, Sara},
  booktitle={Proceedings of the SIGCHI conference on Human factors in computing systems},
  pages={1455--1464},
  year={2007}
}

@article{palinscar1984reciprocal,
  title={Reciprocal teaching of comprehension-fostering and comprehension-monitoring activities},
  author={Palinscar, Aannemarie Sullivan and Brown, Ann L},
  journal={Cognition and instruction},
  volume={1},
  number={2},
  pages={117--175},
  year={1984},
  publisher={Taylor \& Francis}
}

@article{narciss2025learning,
  title={Learning from errors and failure in educational contexts: New insights and future directions for research and practice},
  author={Narciss, Susanne and Alemdag, Ecenaz},
  journal={British Journal of Educational Psychology},
  volume={95},
  number={1},
  pages={197--218},
  year={2025},
  publisher={Wiley Online Library}
}

@article{choudhuri2025utilize,
  title={Insights from the Frontline: GenAI Utilization Among Software Engineering Students},
  author={Choudhuri,Rudrajit and Ramakrishnan, Ambareesh and Chatterjee, Amreeta and Trinkenreich, Bianca and others},
  journal={IEEE Xplore},
  pages={1-12},
  year={2025},
}

@article{zi2025struggle,
  title={“I Would Have Written My Code Differently”: Beginners Struggle to Understand LLM-Generated Code},
  author={Zi, Yangtian and Li, Luisa and Guha, Arjun and Anderson, Carolyn Jane and Feldman, Molly Q},
  journal={Proceedings of the 33rd ACM International Conference on the Foundations of Software Engineering (FSE Companion '25)},
  year={2025},
}

@article{nguyen2024misread,
  title={How Beginning Programmers and Code LLMs (Mis)read Each Other},
  author={Nguyen, Sydney and Babe, Hannah McLean and Zi, Yangtian and Guha, Arjun and Anderson, Carolyn Jane and Feldman, Molly Q},
  journal={ Proceedings of the 2024 CHI Conference on Human Factors in Computing Systems (CHI '24)},
  pages={1-26},
  year={2024},
}

@misc{anonymous_2025_17220238,
  author       = {Anonymous, Anonymous},
  title        = {Replication Package for \name{} for
                   K-12 Students Study
                  },
  month        = sep,
  year         = 2026,
  publisher    = {Zenodo},
  doi          = {10.5281/zenodo.20018098},
  url          = {https://zenodo.org/records/20018099},
}

@book{charmaz2006constructing,
	title        = {Constructing grounded theory: A practical guide through qualitative analysis},
	author       = {Charmaz, Kathy},
	year         = 2006,
	publisher    = {sage}
}

@article{mittal2024comprehensive,
  title={A comprehensive review on generative AI for education},
  author={Mittal, Uday and Sai, Siva and Chamola, Vinay and others},
  journal={IEEE Access},
  year={2024},
  publisher={IEEE}
}

@inproceedings{gu2025ai,
  title={AI literacy in K-12 and higher education in the wake of generative AI: An integrative review},
  author={Gu, Xingjian and Ericson, Barbara J},
  booktitle={Proceedings of the 2025 ACM Conference on International Computing Education Research V. 1},
  pages={125--140},
  year={2025}
}

@article{klopfer2024generative,
  title={Generative AI and K-12 education: An MIT perspective},
  author={Klopfer, Eric and Reich, Justin and Abelson, Hal and Breazeal, Cynthia},
  year={2024},
  publisher={MIT}
}

@inproceedings{alshaikh2020socratic,
  title={A Socratic tutor for source code comprehension},
  author={Alshaikh, Zeyad and Tamang, Lasagn and Rus, Vasile},
  booktitle={International conference on artificial intelligence in education},
  pages={15--19},
  year={2020},
  organization={Springer}
}

@online{britannica_socratic_method,
  author       = {{Encyclopaedia Britannica}},
  title        = {Socratic method},
  year         = {2026},
  url          = {https://www.britannica.com/topic/Socratic-method},
  note         = {Last updated March 13, 2026. Accessed April 15, 2026}
}

@inproceedings{alshaikh2020experiments,
  title={Experiments with a socratic intelligent tutoring system for source code understanding},
  author={Alshaikh, Zeyad and Tamang, Lasang Jimba and Rus, Vasile},
  booktitle={The Thirty-Third International Florida Artificial Intelligence Research Society Conference (FLAIRS-32)},
  year={2020}
}

@inproceedings{al2023socratic,
  title={Socratic questioning of novice debuggers: A benchmark dataset and preliminary evaluations},
  author={Al-Hossami, Erfan and Bunescu, Razvan and Teehan, Ryan and Powell, Laurel and Mahajan, Khyati and Dorodchi, Mohsen},
  booktitle={Proceedings of the 18th Workshop on Innovative Use of NLP for Building Educational Applications (BEA 2023)},
  pages={709--726},
  year={2023}
}

\end{sloppypar}
\end{document}